\documentclass[twocolumn,showpacs,preprintnumbers,amsmath,amssymb]{revtex4}

\usepackage{graphicx}% Include figure files
\usepackage{dcolumn}% Align table columns on decimal point
\usepackage{bm}% bold math

\begin{document}

\title{Close-Packing of Clusters: Application to Al$_{\bf 100}$}

\author{K. Manninen$^1$, J. Akola$^2$, and M. Manninen$^1$}
\affiliation{$^1$Department of Physics, University of Jyv\"askyl\"a,
FIN-40351 Jyv\"askyl\"a, Finland}
\affiliation{$^2$Institut f\"ur Festk\"orperforschung, Forschungszentrum J\"ulich,
D-52428 J\"ulich, Germany}

\date{\today}

\begin{abstract}
The lowest energy configurations of close-packed clusters up to $N=110$ atoms with 
stacking faults are studied using the Monte Carlo method with a Metropolis 
algorithm. Two types of contact interactions, a pair-potential and a many-atom 
interaction, are used. Enhanced stability is shown for $N=12$, 26, 38, 50, 
59, 61, 68, 75, 79, 86, 100 and 102, of which only the sizes 38, 75, 79, 86, and 
102 are pure FCC clusters, the others having stacking faults. A connection between 
the model potential and density functional calculations is studied in the case of 
Al$_{100}$. The density functional calculations are consistent with the experimental 
fact that there exist epitaxially grown FCC clusters starting from relatively small 
cluster sizes. Calculations also show that several other close-packed motifs exist
with comparable total energies. 
\end{abstract}

\maketitle

%\pacs{PACS: 67.40.Db, 73.21.La}

\section{Introduction}

Formation of atomic clusters can lead to close-packed (CP) structures
under some conditions. The shape of such clusters is determined by the balance 
of surface area, surface energy, and internal strain. In small metal 
clusters also the self-deformation of the valence electron density can play an 
important role. Other common packing patterns in clusters include often icosahedral 
and decahedral motifs with internal twin boundaries.\cite{martin96} Metals such 
as Na, Mg and Cu show icosahedral magic numbers.\cite{martin96,reinhard97} 
The motivation to study FCC and other close-packed clusters is the fact that 
some elements, for example Al, are found to form FCC clusters of octahedral 
shape.\cite{martin92} Moreover, at certain cluster sizes pair potentials 
(e.g. Lennard-Jones) also yield FCC-based structures as the most stable 
isomers.

Generally, the search for the lowest energy isomer of a cluster is a difficult 
problem due to the vast number of isomers which correspond to local minima
on a complex potential energy surface. The energy differences between 
the isomers are caused by the surface energy,\cite{Wul01} strain energy due to 
structural defects such as twin boundaries,\cite{Ino69} and in metals also the 
electronic shell structure.\cite{hakkinen97} For example, if one uses classical 
or {\it ab initio} molecular dynamics (MD) one usually needs an appropriate 
initial configuration in order to save computation time. 
\cite{hakkinen95,rytkonen00,kumar91,delaly92,akola99} One purpose of this 
work is to apply the Monte Carlo method to look for the most stable isomers
of hard sphere clusters, to be used as starting geometries
for MD simulations.  

Classical molecular dynamics have been extensively used to study the
lowest energy structures of small clusters. However, finding the correct 
ground state geometry of a small cluster ($N\le 100$) is a difficult 
optimization task even with classical pair potentials.\cite{Wil85} For 
historical reasons, the Lennard-Jones potential is the best 
studied,\cite{Hoa83,doye98,doye99,leary99} but  
other pair potentials\cite{doye01b,doye00,doye01} as well as
many-atom potentials\cite{baletto00,baletto02,valkealahti92,cleveland97} 
have been used. Nearly all give an icosahedral geometry for the 13 and 55 atom 
clusters. The behavior of the binding energy as a function of cluster size is, 
however, quite different for different potentials. While $N=13$ is seen as an
exceptionally stable size, the second complete icosahedron ($N=55$) is not a 
clear local minimum for most of the potentials studied. For example, the 38 
atom FCC structure (Wulff's polyhedron) depends less on the model potential 
than does the 55 icosahedron.

In this article, we present results for the lowest energy isomers 
of close-packed clusters in the size range $N\leq 110$. Our model, which
uses the hard sphere packing as a starting point, does not include icosahedral 
and decahedral motifs as well as any other structures with varying bond lengths 
and angles. This makes our approach extremely efficient in finding
lowest energy structures of CP clusters.
The cluster energy is determined using either a pairwise 
nearest neighbor interaction or a many-atom potential based on the
tight-binding model. For many sizes, the lowest energy structure found  
includes stacking faults, making the cluster a mixture of FCC and HCP phases.

For a cluster with 100 aluminum atoms we use a density functional (DF) method 
to relax the atomic positions of the low energy isomers obtained with the 
simple model. 
Our goal is to study the applicability of our classical energy 
expression in a realistic cluster, where the true electronic 
structure is present.
The cluster size $N=100$ is particularly interesting because
the experimental photoelectron spectrum of the anion has a large energy
gap between the highest and lower-lying occupied orbitals.\cite{Wang98} Our 
calculations show that many of the low energy isomers have such an energy 
gap, but the absolute value of the theoretical gap is still smaller than the 
experimental result.

The plan of this article is the following. In Section 2, the theoretical hard 
sphere model and the DF method used are outlined. The results for the lowest 
energy structures of the close-packed clusters and their relations to the 
cluster shapes are presented in Section 3, and the DF calculations for Al$_{100}$ 
and the corresponding photoelectron spectra are presented in Section 4. The 
conclusions are given in Section 5.

\section{Simulation methods}

\subsection{Monte Carlo method for hard sphere clusters} 

The lowest energy geometries of hard sphere clusters are computed using 
the Monte Carlo (MC) method with a Metropolis algorithm. This is described in 
our earlier work where the role of stacking faults in small hard sphere 
clusters was studied.\cite{kmanninen02} Similar algorithm was earlier 
used by Akola {\it et al.} for making pure FCC clusters.\cite{akola00} 
A dense lattice is generated inside a spherical volume in such a way 
that atoms occupying these lattice sites can form an FCC or HCP lattice 
or any combination of these two, including stacking faults in all possible 
directions. The lattice sites are then randomly populated with $N$ atoms, 
such that the minimum distance between any atom pair is twice the hard 
sphere radius, i.e., the nearest neighbor distance in the FCC lattice. 
After this, a Monte Carlo procedure is used together with the cluster 
binding energy to change the lattice site occupations leading to a 
``clustering'' of atoms. The simulation is started at a high temperature 
that is gradually decreased to zero to obtain the low energy isomers. 
This optimization procedure, including millions of steps, is repeated typically 
at least 1000 times for each cluster size. During the optimization, we record 
not only the most stable geometry but also many other low 
energy isomers.

Two simple models, a pair-potential (PP) and a tight-binding based
potential (TB), are used as a contact interaction between the hard spheres. 
In the former the energy is calculated as
\begin{equation}
\label{pp}
E_{\rm PP}=-{V\over 2} \sum_i^N C_i=-VN_{\rm bonds}
\end{equation}
where $C_i$ is the coordination number of the atom $i$ and
$V$ is the strength of the interaction ($V$ determines the 
energy scale) and $N_{\rm bonds}$ is the total number of bonds
(contacts between the hard spheres). The second model is derived 
from tight-binding theory,\cite{gupta81,finnis84} and
it has a simple square root dependence on the coordination number
\begin{equation}
\label{tb}
E_{\rm TB}=-{V\over 2} \sum_i^N \sqrt{C_i}.
\end{equation}
In practice, Eq.(\ref{tb}) has shown to be a good approximation for 
the true TB energy of small clusters.\cite{hakkinen91} 
Nevertheless, this model cannot describe effects related to the details
of the electronic structure, such as Jahn-Teller deformation, and the
geometries obtained from Eq.(\ref{tb}) should not be confused with the 
most stable geometries determined with the true tight-binding 
method.\cite{wang87,yoshida94}

The Monte Carlo simulations are performed using the PP model energy 
expression Eq.(\ref{pp}), which leads in many cases to several different 
geometries with the same energy. The TB interaction is more practicable 
here, since it is more sensitive in separating the energy of different 
isomers. In most cases, it removes the degeneracy of lowest energy 
isomers of the PP model. The TB model favors geometries where each atom has 
a similar coordination, whereas the PP model is insensitive to the bond 
distribution.

\subsection{Electronic structure calculations}

The DF calculations of Al$_{100}$ isomers are 
performed using the Car-Parrinello molecular dynamics (CPMD) code,\cite{CPMD} 
where the electron-ion interaction is described by an ionic pseudopotential, 
\cite{TM91} and the generalized gradient correction approximation 
of Perdew, Burke, and Ernzerhof (PBE) is applied to the exchange 
correlation energy of the electron density.\cite{PBE96} The basis set 
is taken to be plane waves with a cut-off energy of 15.4 Ry. In contrast to 
many Car-Parrinello simulations, we do not enforce periodicity in the system, 
i.e., calculations are performed in an isolated cubic box of 25.4 {\AA}. 
We also do not employ the Car-Parrinello algorithm for the coupling of 
ionic and electronic solutions during geometry optimization, and the 
electronic Hamiltonian is rediagonalized after each geometry optimization 
step. The metal clusters studied show systematically small energy gaps between 
the occupied and unoccupied molecular orbitals (HOMO-LUMO gaps). In order 
to converge the electron density, a finite temperature functional 
($T=300$ K) is used for the the Kohn-Sham (KS) orbital occupancies. The ionic 
positions are optimized according to a conjugate gradient method until 
all the nuclear gradient components are below $1\times 10^{-4}$ au.

\section{Results}

\subsection{Lowest energy geometries of the hard sphere clusters}

\begin{figure}
\includegraphics{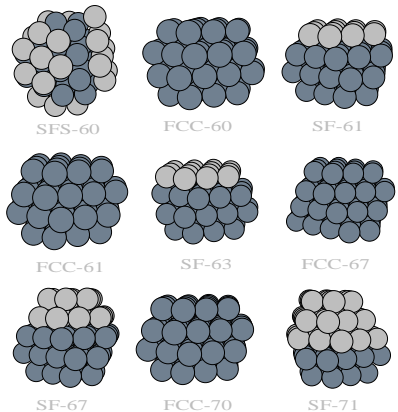}
\caption{Selection of low energy isomers in the size range $N=60-71$.
The different shading of layers illustrates changes in packing (stacking
faults).}
\label{kuva1a} 
\end{figure}

\begin{figure}
\includegraphics{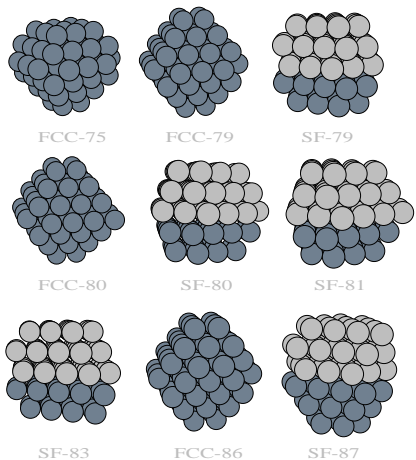}
\caption{Selection of low energy isomers in the size range $N=72-87$.}
\label{kuva1b} 
\end{figure}

\begin{figure}
\includegraphics{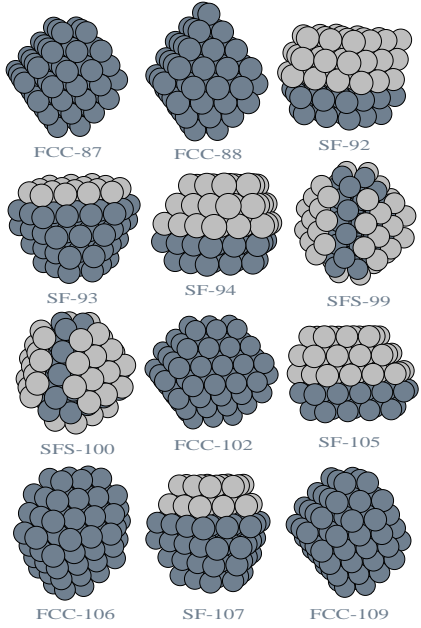}
\caption{Selection of low energy isomers in the size range $N=87-109$.}
\label{kuva1c} 
\end{figure}

Tables \ref{table1}, \ref{table2}, \ref{table3}, and \ref{table4} give 
the energies of the lowest energy isomers found with the PP and TB potentials. 
For each size, we show the energy of the most stable FCC isomer together with 
the lowest energy isomer with one or more stacking faults (SF). In addition, 
the occupation 
numbers of parallel (111) layers in the FCC isomer are also shown (this is 
not done for the SF clusters, since these cannot generally be described by 
parallel (111) layers). Examples of the isomers obtained are shown in 
Figures \ref{kuva1a}, \ref{kuva1b}, and \ref{kuva1c}. The results for the 
most stable isomers of small clusters $N=4-58$ agree with those published earlier, 
\cite{kmanninen02} except for some TB energies. The lowest energy 
structures obtained using the PP model are similar to the results by Doye and 
Wales \cite{doye95} except for $N=33$, 49, 50, 51, 68, 69, 82, 107, 
and 108, where we have found a more stable geometry (for 33, 49, 51, 69, 82, 107 
our results have one additional bond, for 50, 68, 108 two bonds more than 
those of Doye and Wales). 

%%%%%%%%%%%%%%%%%%%%%%%% TABLE 1 %%%%%%%%%%%%%%%%%
\squeezetable
\begin{table}
\caption{Lowest energy FCC and SF isomers in the size range $N=4-40$. For 
the PP model the (negative) energy is given as a number of bonds ($V=1$, see 
Eq.(1)). The same scaling is used for the TB potential. The column 'layers' 
gives the number of atoms on each close-packed (111) layer for the most stable
FCC structure. For $N=4$, 38, 39, and 40 there is no SF structure with 
the same number or more bonds than in the most stable FCC structure.}
\label{table1}
\begin{center}
\vspace{2pt}
\begin{tabular}{r r r c r r}  \hline

$N$ & $-E_{\rm PP}^{\rm FCC}$ & $-E_{\rm PP}^{\rm SF}$ & layers  &  
$E_{\rm TB}^{\rm FCC}$ & $E_{\rm TB}^{\rm SF}$\\  \hline

4 & {\bf 6} &  & (3,1) & {\bf -3.464} &  \\
5 & 8 & {\bf 9} & (4,1) & -4.439  & {\bf -4.732}  \\
6 & 12 & 12 & (3,3) & {\bf -6.000} & -5.968   \\
7 & 15 & 15 & (4,3) & {\bf -7.220} & -7.180  \\
8 & 18 & 18 & (5,3) & -8.429 & {\bf -8.434}  \\
9 & 21 & 21 & (5,4) & -9.650 & -9.650  \\
10 & 25 & 25 & (5,5) & {\bf -11.118} & -11.107   \\ 
11 & 28 & {\bf 29} & (6,5) & -12.327 & {\bf -12.532}   \\
12 & 32 & {\bf 33} & (7,5) & -13.745 & {\bf -13.951}  \\
13 & 36 & 36 & (7,6) & {\bf -15.177} & -15.148   \\ 
14 & 40 & 40 & (4,7,3) & -16.575 & {\bf -16.581}  \\
15 & 44 & 44 & (4,7,4) & {\bf -18.002} & -18.000  \\
16 & 48 & 48 & (6,7,3) & -19.403 & {\bf -19.420}  \\
17 & 52 & 52 & (5,7,5) & -20.822 & -20.822  \\
18 & 56 & 56 &  (6,7,5) & -22.214 & {\bf -22.240}   \\
19 & 60 & 60 & (7,8,4) & -23.643 & -23.643   \\
20 & 64 & 64 & (7,8,5) & -25.053 & -25.053  \\
21 & 68 & 68 & (8,8,5) & -26.454 & {\bf -26.460} \\
22 & 72 & 72 & (8,8,6) & -27.852 & {\bf -27.863}  \\
23 & 76 & 76 & (7,10,6) & -29.262 & {\bf -29.276}  \\
24 & 81 & 81 & (7,10,7) & {\bf -30.869} & -30.847  \\
25 & 85 & 85 & (8,10,7) & -32.270 & -32.270 \\
26 & 89 & {\bf 90} & (9,10,7) & -33.671 & {\bf -33.844}  \\
27 & 93 & {\bf 94} & (9,10,8) & -35.064 & {\bf -35.254}  \\
28 & 97 & {\bf 98} & (8,12,8) & -36.471 & {\bf -36.671}  \\
29 & 102 & 102 & (10,12,7) & -38.070 & {\bf -38.075} \\
30 & 106 & 106 & (11,12,7) & -39.471 & {\bf -39.477}  \\
31 & 111 & 111 & (12,12,7) & {\bf -41.070} & -41.065  \\
32 & 115 & 115 & (12,12,8) & {\bf -42.462} & -42.458  \\
33 & 119 & {\bf 120} & (12,12,9) &  -43.855 & {\bf -44.045} \\
34 & 124 & 124 & (9,12,9,4) &  -45.434 & {\bf -45.441} \\ 
35 & 128 & {\bf 129} &  (7,12,11,5) &  -46.816 & {\bf -47.045} \\ 
36 & 133 & 133 & (7,12,12,5) & -48.410 & {\bf -48.437} \\
37 & 138 & 138 & (7,12,11,7) & -49.996 & {\bf -50.003} \\
38 & {\bf 144} &  & (7,12,12,7) & {\bf -51.786} &  \\
39 & {\bf 148} &  & (8,12,12,7) & {\bf -53.179} &  \\
40 & {\bf 152} &  & (9,12,12,7) & {\bf -54.571} &   \\ 
\hline
\end{tabular}
\end{center}
\end{table}

%%%%%%%%%%%%%%%%%%% Fig. 1 %%%%%%%%%%%%%%

%%%%%%%%%%%%%%%%%%% Fig. 2 %%%%%%%%%%%%%%

%%%%%%%%%%%%%%%%%%% Fig. 3 %%%%%%%%%%%%%%

The total number of bonds in the most stable geometry is the same in both the 
PP and TB models for all $N$. The difference between the two energy 
formulas appears only in separation between different isomers with the same 
number of bonds. For each cluster size we have determined the lowest energy 
FCC geometry, but the most stable geometry with at least one stacking fault is 
determined only for those clusters where a stacking fault does not decrease 
the number of bonds (as in the sizes 4, 38, 39, 40, 86, 88, 102, 104). For 
the other sizes (see Tables I-IV), the TB model gives lower energies for  
SF structures in many cases. A detailed discussion of the structure evolution 
of small hard sphere clusters has been given earlier,\cite{kmanninen02} and
we concentrate on clusters with more than 60 atoms in the following. 

%%%%%%%%%%%%%%%%%%%%%%%% TABLE 2 %%%%%%%%%%%%%%%%%

\begin{table}
\caption{As in Table I, but for $N=41-60$.}
\label{table2}
\begin{center}
\vspace{2pt}
\begin{tabular}{r r r c r r}  \hline
$N$ & $-E_{\rm PP}^{\rm FCC}$ & $-E_{\rm PP}^{\rm SF}$ & layers  &  $E_{\rm TB}^{\rm FCC}$ & $E_{\rm TB}^{\rm SF}$  \\  \hline
 41 & 156 & 156 & (8,14,12,7) & -55.964 & {\bf -55.974} \\
 42 & 160 & 160 & (8,14,13,7) & -57.360 & {\bf -57.388} \\
 43 & 165 & 165 & (8,14,13,8) & -58.961 & {\bf -58.972} \\
 44 & 169 & 169 & (11,14,12,7) & -60.349 & {\bf -60.365} \\
 45 & 174 & 174 & (10,14,13,8) & -61.953 & {\bf -61.964} \\
 46 & 178 & 178 & (11,14,13,8) & -63.345 & {\bf -63.350} \\
 47 & 183 & 183 & (12,14,13,8) & {\bf -64.936} & -64.933 \\
 48 & 187 & 187 & (10,16,14,8) & -66.308 & {\bf -66.342} \\
 49 & 191 & {\bf 192} & (10,16,14,9) & -67.701 & {\bf -67.933} \\
 50 & 196 & {\bf 198} & (10,16,16,8) & -69.284 & {\bf -69.663}  \\
 51 & 201 & {\bf 202} & (10,16,16,9) & -70.870 & {\bf -71.056} \\
 52 & 207 & 207 & (10,16,16,10) & {\bf -72.652} & -72.638 \\
 53 & 211 & 211 & (11,16,16,10) & {\bf -74.044} & -74.023 \\
 54 & 216 & 216 & (12,16,16,10) & {\bf -75.635} & -75.625 \\
 55 & 220 & 220 & (12,16,16,11) & {\bf -77.027} & -77.004 \\
 56 & 225 & 225 & (12,16,16,12) & {\bf -78.617} & -78.608 \\
 57 & 229 & 229 & (13,16,16,12) & {\bf -80.010} & -79.987 \\ 
 58 & 233 & {\bf 234} & (13,19,16,10) & -81.360 & {\bf -81.587} \\ 
 59 & 238 & {\bf 240} & (12,18,17,12) & -82.977 & {\bf -83.272} \\ 
 60 & 243 & {\bf 244} & (12,18,19,11) & -84.535 & {\bf -84.651} \\
\hline
\end{tabular}
\end{center}
\end{table}

%%%%%%%%%%%%%%%%%%%%%%%% TABLE 3 %%%%%%%%%%%%%%%%%

\begin{table}
\caption{As in Table I, but for $N=61-80$.}
\label{table3}
\begin{center}
\vspace{2pt}
\begin{tabular}{r r r c r r}  \hline
$N$ & $-E_{\rm PP}^{\rm FCC}$ & $-E_{\rm PP}^{\rm SF}$ & layers  &  $E_{\rm TB}^{\rm FCC}$ & $E_{\rm TB}^{\rm SF}$  \\  \hline
 61 & 249 & 249 & (12,19,18,12) & -86.316 & {\bf -86.333} \\
 62 & 253 & 253 & (12,19,18,13) & -87.709 & {\bf -87.726} \\
 63 & 258 & 258 & (14,19,18,12) & -89.299 & {\bf -89.306} \\
 64 & 262 & 262 & (13,18,19,14) & -90.692 & {\bf -90.698} \\
 65 & 267 & 267 & (12,18,19,16) & -92.285 & -92.285 \\
 66 & 271 & 271 & (13,18,19,16) & -93.675 & {\bf -93.681} \\
 67 & 276 & 276 & (12,18,19,18) & {\bf -95.265} & -95.232 \\
 68 & 282 & 282 & (7,12,18,19,12) & -96.981 & {\bf -96.998} \\
 69 & 286 & 286 & (12,19,18,13,7) & -98.360 & {\bf -98.390} \\
 70 & 291 & 291 & (14,21,21,14) & -99.981 & -99.981 \\
 71 & 296 & 296 & (9,16,21,16,9) & -101.542 & {\bf -101.548} \\
 72 & 300 & 300 & (16,21,21,14) & {\bf -102.964} & -102.953 \\
 73 & 305 & 305 & (12,18,19,15,9) & -104.501 & {\bf -104.514} \\
 74 & 310 & 310 & (10,16,18,18,12) & {\bf -106.087} & -106.085 \\
 75 & 316 & 316 & (12,18,19,16,10) & {\bf -107.860} & -107.856\\
 76 & 320 & 320 & (11,16,19,18,12) & {\bf -109.252} & -109.248 \\
 77 & 325 & 325 & (12,18,19,18,10) & {\bf -110.819} & -110.818 \\ 
 78 & 330 & 330 & (11,18,19,18,12) & {\bf -112.396} & -112.387 \\ 
 79 & 336 & 336 & (12,18,19,18,12) & {\bf -114.177} & -114.163 \\ 
 80 & 340 & 340 & (12,18,19,18,13) & {\bf -115.570} & -115.528 \\
\hline
\end{tabular}
\end{center}
\end{table}

Clusters with $N=58$, 59, and 60 are based on a truncated 31-atom tetrahedron 
with all four overlayers in stacking fault locations. We denote such 
isomers with tetrahedral symmetry as SFS (see also $N=100$). Both the 
PP and TB models give a large energy difference between the most
stable FCC geometry and the SFS lowest energy isomer. For $62\leq N\leq 64$, 
the clusters consist mainly of four (111) layers, three of them being in an 
FCC arrangement and the fourth being either FCC or HCP. For each size, the most 
stable FCC and SF structures have the same number of atoms in the layers. 
The lowest energy SF structure of $N=67$ resembles closely an HCP cluster. 
It consists of five layers, four of them forming an HCP-lattice and the
fifth in a stacking fault position (layer-packing ABABC, which can be seen
as two connected FCC subunits). SF clusters with $N=68$, 69, and 73  
have similar structures, whereas the cluster sizes 70, 72 and 74 atoms show 
four parallel FCC layers, and the fifth layer is displaced in a stacking 
fault position. The lowest energy structure of 71 atoms consists
of two FCC subdomains (layer-packing ABCBA).

In the size range 75--80, the lowest energy isomers are FCC structures according 
to the TB potential with a very small energy difference to SF structures. 
This applies also for $N=79$, where the FCC construction gives a complete 
truncated octahedron (TO). The corresponding SF isomer (two FCC units connected) 
also has large (111) facets, which explains the low energy. Similar kinds 
of SF clusters with a stacking fault layer inside the cluster are found in 
$N=81-92$, and 94, while $N=81$, 83, and 92 are better than 
FCC for both potentials. The clusters with 86 and 88 atoms have a nondegenerate 
FCC energy minimum in the PP model. These isomers are based on the FCC-79 with 
one (111) overlayer. For $N=87$, the FCC structure is still lower 
in TB energy, but there are SF structures with the same number of bonds (PP 
energy). The lowest energy isomers of $N=93$ and 95 consist of four
FCC layers, with the fifth being in HCP position (with respect to the two 
lower layers). The clusters with  $N=96-98$ atoms do not have stacking
faults according to the TB model.

%%%%%%%%%%%%%%%%%%%%%%%% TABLE 4 %%%%%%%%%%%%%%%%%

\begin{table}
\caption{As in Table I, but for $N=81-110$. For $N=86$, 88, 102, and 104 
there is no SF structure with the same number or more bonds than in the 
most stable FCC structure.}
\label{table4}
\begin{center}
\vspace{2pt}
\begin{tabular}{r r r c r r}  \hline
$N$ & $-E_{\rm PP}^{\rm FCC}$ & $-E_{\rm PP}^{\rm SF}$ & layers  &  $E_{\rm TB}^{\rm FCC}$ & $E_{\rm TB}^{\rm SF}$  \\  
\hline
 81 & 344 & {\bf 345} & (13,18,19,18,13) & -116.962 & {\bf -117.126} \\
 82 & 349 & 349 & (10,16,21,21,14) & -118.524 & {\bf -118.535} \\
 83 & 353 & {\bf 354} & (13,18,21,19,12) & -119.917 & {\bf -120.126} \\
 84 & 358 & 358 & (12,21,21,18,12) & -121.483 & {\bf -121.518} \\
 85 & 363 & 363 & (12,18,21,20,14) & -123.069 & {\bf -123.108} \\
 86 & {\bf 369} &  & (12,18,21,21,14) & {\bf -124.842}  \\
 87 & 373 & 373 & (15,21,21,18,12) & {\bf -126.235} & -126.221 \\
 88 & {\bf 378} &  & (13,19,21,21,14) & {\bf -127.825} & \\
 89 & 382 & 382 & (16,21,21,18,13) & {\bf -129.218} & -129.204 \\
 90 & 387 & 387 & (12,19,24,21,14) & -130.741 & {\bf -130.773} \\
 91 & 391 & 391 & (12,18,23,23,15) & -132.165 & {\bf -132.169} \\
 92 & 396 & {\bf 397} & (12,19,24,22,15) & -133.717 & {\bf -133.902} \\
 93 & 402 & 402 & (12,18,23,24,16) & -135.507 & {\bf -135.529} \\
 94 & 407 & 407 & (16,23,24,19,12) & -137.067 & {\bf -137.069} \\
 95 & 411 & 411 & (13,19,23,24,16) & -138.490 & {\bf -138.502} \\
 96 & 416 & 416 & (19,24,23,18,12) & {\bf -140.069} & -140.052 \\
 97 & 421 & 421 & (16,23,24,21,13) & {\bf -141.621} & -141.611 \\ 
 98 & 427 & 427 & (16,23,24,21,14) & {\bf -143.385} & -143.377 \\ 
 99 & 431 & {\bf 432} & (16,23,24,21,15) & -144.764 & {\bf -144.846} \\ 
 100 & 436 & {\bf 438} & (16,23,24,24,15) & -146.362 & {\bf -146.614} \\
 101 & 441 & 441 & (16,23,24,23,15) & {\bf -147.930} & -147.754 \\
 102 & {\bf 447} &  & (16,23,24,23,16) & {\bf -149.703} &  \\
 103 & 451 & 451 & (16,23,24,23,17) & {\bf -151.096} & -151.052\\
 104 & {\bf 456} &  & (18,23,24,23,16) & {\bf -152.686} &   \\
 105 & 460 & {\bf 461} & (18,26.26,21,14) & -154.050 & {\bf -154.245} \\
 106 & 465 & 465 & (16,24,27,23,16) & {\bf -155.611} & -155.578 \\
 107 & 470 & 470 & (19,27,27,21,13) & -157.164 & {\bf -157.169} \\
 108 & 476 & 476 & (19,27,27,21,14) & -158.928 & {\bf -158.934} \\ 
 109 & 480 & 480 & (16,23,26,26,18) & {\bf -160.368} & -160.315 \\ 
 110 & 485 & 485 & (16,24,17,25,18) & {\bf-161.928} & -161.917 \\ 
\hline
\end{tabular}
\end{center}
\end{table}

The most stable isomers for 99 or 100 atoms have an SFS structure based 
on a truncated FCC tetrahedron with stacking faults at each of the four surfaces. 
The 100 atom cluster has the same structural motif as SFS-59, whereas in the 
case of 99 (101) atoms, one surface atom is removed (added). The lowest-lying 
TB geometries for $101\leq N\leq 104$, 
$N=106$, 109, and 110 are FCC structures with five (111) layers. They 
all are related to FCC-102, which is an elongated TO having a full atom shell 
(similar to FCC-52). Clusters with $N=105$, 107 and 108 prefer stacking faults 
in the TB model, and SF-105 has the largest number of bonds in its class.

\subsection{Magic numbers}

\begin{figure}
\includegraphics{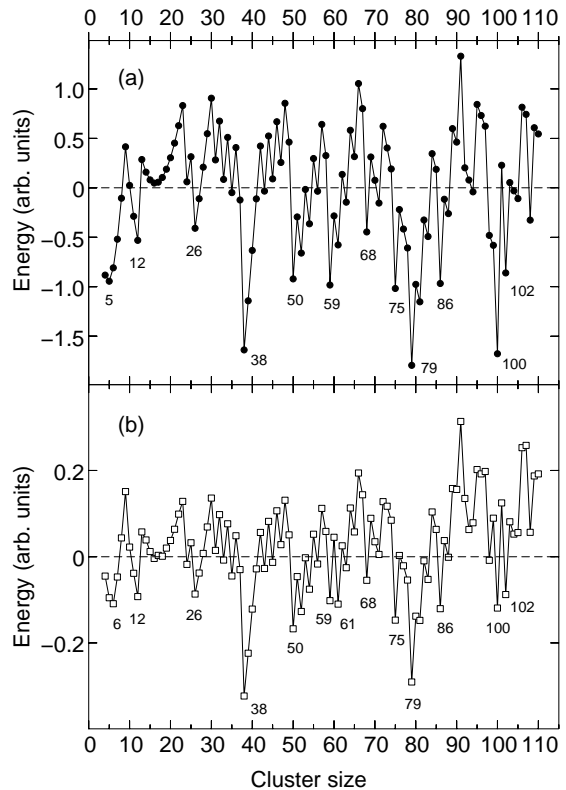}
\caption{Deviation of the lowest energy of (a) PP and (b) TB models from 
that calculated using the mass formula: $E_{\rm ave} = aN+bN^{2/3}+cN^{1/3}$. 
We fix the first coefficient to the bulk value $a=-6$ ($a=-\sqrt{3}$) and obtain 
the best fit with $b=7.651$ and $c=-0.250$ ($b=1.160$ and $c=0.369$).}
\label{enero}
\end{figure}

The most stable cluster sizes (magic numbers) reflect the stability 
of the cluster with respect to the neighboring sizes. In order to see also 
the possible regions of increased stability, it is convenient to subtract 
a smooth size dependence from the energy. This can be obtained by fitting 
the total energies to a ``mass formula''
$E_{\rm ave}=-E_{\rm coh}N+bN^{2/3}+cN^{1/3}$, where $b$ and $c$ are fitting
parameters and $E_{\rm coh}$ is the cohesion energy of the model
in question ($E_{\rm coh}=-6V$ for the PP model and $-V\sqrt{3}$ for the 
TB model).\cite{Wei35,Per91} Figure \ref{enero} shows the deviation of 
energy from this function. Clusters with $N=12$, 26, 38, 50, 59, 61, 68, 
75, 79, 86, 100, and 102 are the most pronounced local minima, but there 
are several weaker local minima, and the results show odd-even alternation
in some regions. However, since this behavior is not of electronic 
origin,\cite{manninen94} the minima can be for $N$ even
(around 38) or odd (around 61).

%%%%%%%%%%%%%%%%%%%%  Fig. 4 %%%%%%%%%%%%%%%%%%%%%%%%%

The general profile of the energy curve is related to the total number of 
bonds. This is natural in our model which totally neglects the internal strain.
In all cases, the lowest PP and TB structures have the same number of 
bonds and qualitatively similar energy curves. Note that from the magic 
sizes only those with 38, 75, 79, 86, and 102 atoms are pure FCC clusters, 
while others have stacking faults. Doye and Wales have used a pair-wise Morse 
potential in the size range $20\leq N\leq 80$ to study the effect of potential 
range on the magic numbers.\cite{doye97} For the hardest potential studied, 
they found magic numbers 26, 38, 50, 55, 59, 61, 68, and 79, which are  
present in our results except the size 55, which is an icosahedron. 
These results seem to indicate that the internal strain is not important
in clusters which do not have twin boundaries.

\subsection{Moments of inertia}

\begin{figure}
\includegraphics{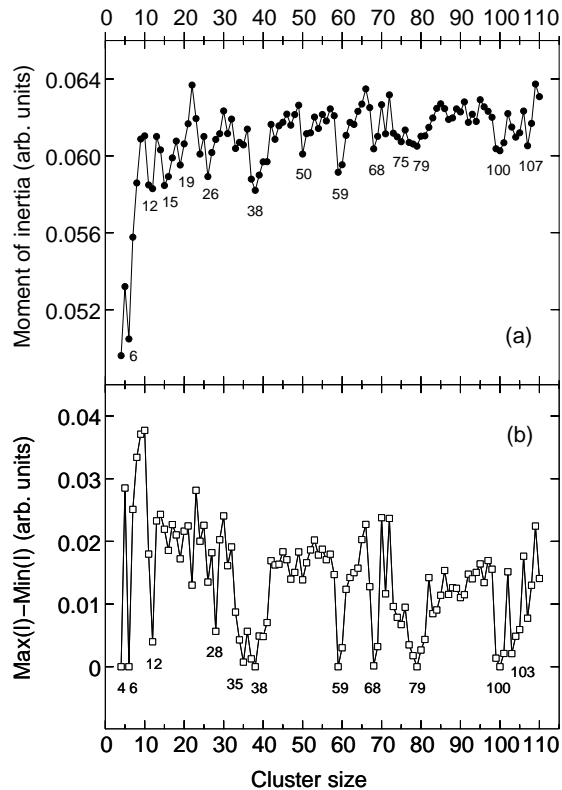}
\caption{Moments of inertia as a function of the cluster size. (a) the normalized
average moment of inertia, and (b) the difference between the largest and smallest 
principal moment of inertia divided by the average value.
}
\label{moment}
\end{figure}

The overall shape of clusters is studied by calculating the three moments 
of inertia for the principal axis, and the normalized average moment of 
inertia
\begin{equation}
\label{mi}
I={{1}\over{N^{5/3}}}\sum_i^N ({\bf R}_i-{\bf R}_{\rm cm})^2,
\end{equation}
where ${\bf R}_i$ is the atom position and ${\bf R}_{\rm cm}$ 
the center of mass (in units of the FCC lattice constant). The factor 
$1/{N^{5/3}}$ is chosen because the moment of inertia is 
proportional to $N^{5/3}$ for a spherical cluster. The normalized moments 
of inertia are shown in Figure \ref{moment}(a), and the similarity between 
Figures \ref{enero} and \ref{moment}(a) is obvious. The minima in moments 
of inertia are present also in the energy curve except for $N=15$, 19, 
and 107.

%%%%%%%%%%%%%%%%%%%% Fig. 5 %%%%%%%%%%%%%%%%%%%%%

It is also interesting to resolve the cluster deformation on the basis
of moments of inertia, since for a sphere (or a cube, etc.) all the 
principal components of inertia are equal. Figure \ref{moment}(b) shows 
the difference of the maximum and minimum components as a function of 
cluster size. Many of the magic clusters ($N=38$, 59, 68, 79, 100)   
have high symmetry, but some, e.g. 50, 86, and 102 have a marked deformation. 
The large (111) facets compensate the increase in surface area (deformation) 
in such cases.

\subsection{Electronic structure calculations of Al$_{100}$ isomers}

\begin{figure}
\includegraphics{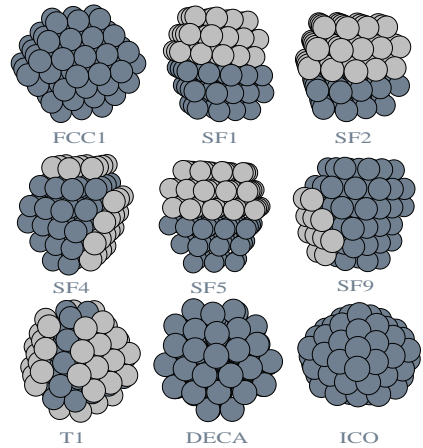}
\caption{Al$_{100}$ isomers and their abbreviations.
}
\label{size100}
\end{figure}

In order to test the applicability of our classical energy expressions 
for Al clusters we have chosen an MC-generated test set of 15 low energy 
isomers for $N=100$ together with the corresponding 
icosahedral and decahedral isomers, which we optimized using the DF method. 
The results for total energies, bond lengths, deformation, 
and HOMO-LUMO gaps ($E_g$) of KS orbitals are presented in 
Table \ref{table5}. The corresponding cluster geometries are shown 
in Figure \ref{size100}. The FCC isomers are all based on the same structural 
motif, which leads to a full atomic shell of an elongated TO 
at cluster size 102 (notice the two missing atoms in Figure \ref{size100}). 
The higher degree of freedom in the case of clusters with stacking faults 
results in a variety of different structures. Among the 11 different geometries 
chosen from this class, five of them (SF2, SF3, SF6, SF7, and SF8) have
the same structure of two connected FCC subdomains with two 
external atoms changing their positions on the surface, and we show 
the most stable geometry SF2. As representatives of CP clusters, the SFS 
isomers T1 and T2 have stacking faults in all four 
[111] directions. The perfect symmetry of T1 is broken in T2 (not shown)
via a surface atom displacement. The icosahedral and decahedral isomer structures 
are obtained from the Cambridge Cluster Database,\cite{walesco} where ICO 
corresponds to the most stable icosahedral configuration found with a model 
potential, and DECA is based on the Mark's decahedron for 101 atoms 
(one atom is removed).

%%%%%%%%%%%%%%%%%%%% Fig. 6 %%%%%%%%%%%%%%%

%%%%%%%%%%%%%%%%%%%%%%%% TABLE 5 %%%%%%%%%%%%%%%%%

\begin{table*}
\caption{Properties of Al$_{100}$ isomers calculated using both the classical
potentials and DF method. Boldfaced letters {\bf o} and {\bf p} refer to oblate 
and prolate deformations, respectively. 
$D$ is a deformation parameter defined in the text.
$\Delta E_{\rm DF}$ is the energy difference between the total energy of the isomer and 
that of the ground state.
Energy differences $\Delta E_{\rm TB}$ 
and $\Delta_{\rm TB}^{\rm DEF}$ are
calculated without and with the deformation correction (Eq.(\ref{tbdef})), and
converted to standard energy units (eV) by taking the binding energy per atom of the 
T1 isomer to be 3.00 eV.
$r$ and $\Delta r$ are the average nearest neighbour distance and its standard deviation,
respectively. $E_g$ is the energy gap between the lowest unoccupied and highest occupied
single electron state (HOMO-LUMO gap).
}
\label{table5}
\begin{center}
\vspace{2pt}
\begin{tabular}{l c c c c c c c}  
\hline
{\rm Isomer} & $-E_{\rm PP}$ & $-E_{\rm TB}$ & D & $\Delta E_{\rm DF}$ (eV) &  $\Delta E_{\rm TB}/\Delta_{\rm TB}^{\rm DEF}$ (eV) & $r/\Delta r$ (\AA) & $E_g$ (eV) \\
\hline
FCC1 & 436 & 146.3443 & {\bf o} 0.256 & 0.000 & 0.552/0.000  & 2.827/0.052 & 0.162  \\
SF1  & 436 & 146.3522 & {\bf o} 0.246 & 0.087 & 0.536/-0.014 & 2.829/0.058 & 0.210  \\
FCC2 & 436 & 146.3615 & {\bf o} 0.246 & 0.129 & 0.517/-0.033 & 2.827/0.055 & 0.067  \\
FCC3 & 436 & 146.3443 & {\bf o} 0.230 & 0.172 & 0.552/0.017  & 2.824/0.055 & 0.080  \\
FCC4 & 435 & 146.1399 & {\bf o} 0.227 & 0.402 & 0.970/0.439  & 2.823/0.058 & 0.099  \\
SF2  & 436 & 146.3304 & {\bf o} 0.242 & 0.656 & 0.580/0.033  & 2.829/0.069 & 0.185  \\
SF3  & 436 & 146.3113 & {\bf o} 0.235 & 0.661 & 0.619/0.078  & 2.827/0.060 & 0.178  \\
SF4  & 435 & 146.1129 & {\bf p} 0.226 & 0.727 & 1.025/0.495  & 2.835/0.073 & 0.140  \\
SF5  & 435 & 146.0905 & {\bf p} 0.222 & 0.766 & 1.071/0.548  & 2.830/0.062 & 0.173  \\
SF6  & 436 & 146.3257 & {\bf o} 0.245 & 0.812 & 0.590/0.041  & 2.830/0.074 & 0.143  \\
SF7  & 436 & 146.3390 & {\bf o} 0.287 & 0.930 & 0.563/0.034  & 2.829/0.062 & 0.111  \\
SF8  & 435 & 146.1521 & {\bf o} 0.255 & 0.993 & 0.945/0.393  & 2.829/0.063 & 0.078  \\
SF9  & 434 & 145.9101 & {\bf p} 0.183 & 1.021 & 1.440/1.018  & 2.832/0.064 & 0.139  \\
T1   & 438 & 146.6140 &         0.000 & 1.043 & 0.000/(1.043)& 2.826/0.048 & 0.022  \\
T2   & 436 & 146.2246 &         0.054 & 1.133 & 0.797/1.238  & 2.828/0.064 & 0.135  \\
...  & ... & ...      &         ...   & ...   & ...          & ...         & ...    \\
DECA &     &          &         0.053 & 2.647 &              & 2.838/0.079 & 0.049  \\
ICO  &     &          & {\bf o} 0.156 & 3.320 &              & 2.851/0.093 & 0.130  \\
\hline
\end{tabular}
\end{center}
\end{table*}

%%%%%%%%%%%%%%%%%%%%%%%%%%%%%%%%%%%%%%%%%%%

\begin{figure}
\includegraphics{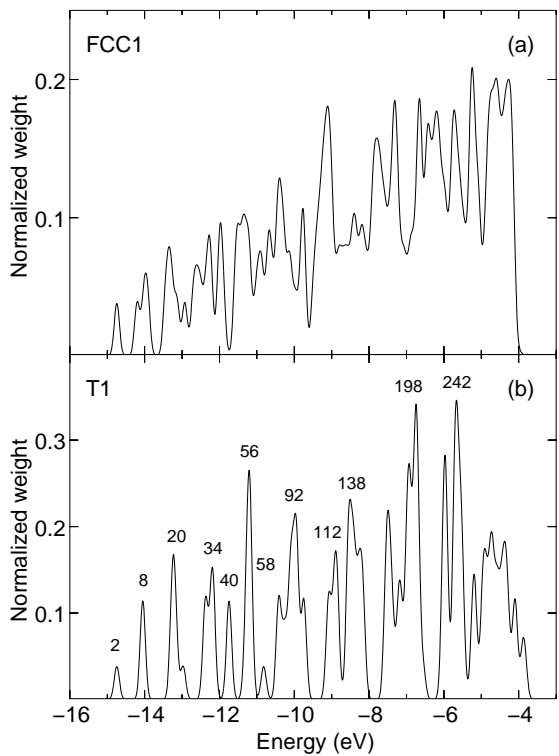}
\caption{DOS of Al$_{100}$ isomers FCC1 and T1. Labeled peaks in the T1 spectrum
refer to the corresponding number of valence electrons (KS orbitals).}
\label{aldos}
\end{figure}

The isomers in Table \ref{table5} are ordered according to the DF total energy. 
The FCC isomers have lowest energy, followed by other CP structures, 
and the DECA and ICO isomers are significantly higher due to the internal strain 
within these geometries. This is in agreement with a previous study of Al clusters, 
\cite{akola00} which showed that Al tends to form FCC geometries at a relatively 
early stage ($N\ge 55$). In contrast to the classical potentials, DF calculations 
give a higher total energy for T1 than for the other CP clusters (except for T2). 
The highly symmetric geometry of T1 results in degeneracies of 
electronic levels (see Figure \ref{aldos}), and one of them occurs at 
the Fermi energy, leading to an energetically unstable situation. The cluster 
undergoes a small Jahn-Teller deformation, which can be seen as a finite 
HOMO-LUMO gap in Table \ref{table5}, but there are no changes in the overall 
shape. We presume that a deformation of the T1 shape should lower the total 
energy, especially when the other structures show marked deformations 
(see Table \ref{table5}). This is related to the self-deformation of valence 
electron density in the jellium model,\cite{Bra93} a phenomenon that lowers total 
energy. Taking this effect into account, we have modified our expression for 
the classical energy:
\begin{equation}
\label{tbdef}
E_{\rm TB}^{\rm DEF}= E_{\rm TB} + {1\over2}K(D-D_0)^2,
\end{equation}
where $K$ is a coupling constant and $D_0$ corresponds to the minimum energy 
deformation,
defined as $D=(I_{\rm max}-I_{\rm min})/I_{\rm ave}$ where $I$'s are the
moments of inertia in the principal axis presentation.
The value of $K$ is calibrated using the DF results for 
FCC1 and T1 (difference in binding energy), assuming that FCC1 represents 
an ideal deformation. The results for the new classical energy in Table \ref{table5}
now correlate better with DF calculations, but SF2 and structures related to it 
remain close to the lowest energy isomer, indicating that other contributions 
(such as surface and strain energy) must also be considered.

Density functional calculations show that the CP isomers lie within a 
very narrow energy range of 1.13 eV (Table \ref{table5}), corresponding 
to 90 K when converted to vibrational energy.\cite{tvib} Moreover, 
the energy difference between the most stable (FCC1) and next higher (SF1) 
isomers is negligible, which emphasizes that one cannot claim that FCC clusters 
are more stable than SF structures. The nearest neighbor distances reveal, 
however, some minor deviations: in FCC clusters the average bond length is 
slightly smaller, and the related distribution width is narrow. The FCC 
isomers are evidently relatively strain-free, which partially explains the 
energetic trend in the CP data set (see also ICO and DECA). As discussed 
above, the electronic structure contributes to the total energy, as shown
by the HOMO-LUMO gap, which is maximized via a self-deformation process whenever 
possible. It is shown in Table \ref{table5} that the HOMO-LUMO gaps of Al$_{100}$ 
clusters are in the range of 0.02-0.21 eV, and that the most stable structures 
FCC1 and SF1 also have significant values. Nevertheless, there is no clear trend 
among the clusters studied. 

%%%%%%%%%%%%%%%%%%%% Fig. 7 %%%%%%%%%%%%%%%

The electronic density of states (DOS) of FCC1 and T1 clusters is shown 
in Figure \ref{aldos}. The DOS of T1 is highly peaked due to the electron 
level degeneracies whereas FCC1 shows a gradually increasing DOS with less 
fine structure. The latter applies basically to all the other CP isomers 
that have no symmetry in the cluster geometry. In order to compare our results 
with the spherical jellium model (SJM), we have labeled the main peaks present 
in the T1 spectrum according to the related number of valence electrons
(KS orbitals). Apart from $N_{el}=56$, the system corresponds fully with the 
magic numbers of SJM up to 198 valence electrons, after which the exact details 
in the cluster shape and structure start to contribute. The T1 cluster actually 
has a $T_d$ symmetry, but a corner truncation has brought it apart from a perfect 
tetrahedron, and no magic numbers related to the tetrahedral external potential 
can be observed.\cite{Rei97} The last electron shell in the T1 spectrum is 
only partially filled, and there is no shell closing at $N_{el}=300$ (Fermi 
energy).

The experimental photoelectron spectrum (PES) of Al$_N$ ($N=100-112$) cluster
anions \cite{Wang98} shows a marked gap at the threshold region. Based on our 
earlier experience with Al clusters,\cite{akola99,akola00} we have compared 
the DOS of close-packed Al$_{100}^-$ isomers (not shown) with the 
experimental PES. As indicated by the sizable HOMO-LUMO gaps of neutral 
clusters, qualitatively correct features can be observed in the DOS of 
lowest energy isomers (FCC1 and SF1). However, the separation of the first 
peak in the theoretical DOS is far too small (0.2 eV), and the corresponding 
electron detachment energy is 3.2 eV, whereas it is 3.4 eV in the experiments. 
This suggests that the experimental spectrum is dominated by an electronically 
stable isomer that is not considered in the present study. Presumably, the long 
thermal tail in the experimental PES (starting from 3.1 eV) is caused by the 
presence of other isomers -- such as FCC1 and SF1.

\section{Conclusion}

Close-packed geometries are important structures present in a small 
cluster size regime, and we have studied the structures of hard sphere 
clusters up to $N=110$ atoms using two types of classical potentials. The total 
energy is minimized using the Monte Carlo method. For most sizes, the discrete
PP model leads to several isomers with the same total energy (number of bonds). 
Only for $N=4$, 38--40, 86, 88, 102, and 104 does the FCC geometry 
have more bonds than any of the geometries with stacking faults. On the other 
hand, for $N=5$, 11, 12, 26--28, 33, 35, 49--51, 58--60, 81, 83, 92, 99--100, 
105 the most stable PP isomer does not have an FCC structure. Clusters 
with $N=58-60$ and $N=99-100$ atoms have a tetrahedral symmetry and stacking 
faults on all surface facets. An inclusion of the TB model yields qualitatively 
same results, the only effect being in the separation of the isomers having 
the maximal coordination.

The energy as a function of the cluster size shows that clusters with $N=12$, 26, 
38, 50, 59, 61, 68, 75, 79, 86, 100, and 102 have the most pronounced energy minima. 
Of these only the 38, 75, 79, 86, and 102 atom clusters have an FCC structure. 
The moments of inertia correlate well with the energy curve, showing that most of 
the magic clusters have also a compact geometry, i.e., the overall shape of clusters 
is not deformed. There exist, however, structures such as FCC-102 where the 
deformation is compensated by the large (111) facets. Such a behavior becomes 
increasingly important as the cluster size increases, leading to an epitaxial 
growth pattern.\cite{Val95}

The connection between the model potential and DF calculations has been studied in 
the case of Al$_{100}$. The DF calculations show that the strain-free CP structures 
are lower in total energy than the corresponding icosahedral and decahedral 
isomers. In CP regime the total energy differences are very small 
(supposing that the coordination number of the cluster is close to the maximum), 
and the electronic structure becomes important. As illustrated by the isomer T1, 
the valence electron density of Al$_{100}$ clusters prefers deformation, and this 
criterion is fulfilled by almost all the CP isomers presented. None of the 
structures reported reproduces the experimental PES.\cite{Wang98} The 
exceptional shape of the experimental curve and the high electron detachment 
energy indicate that the underlying isomer must be electronically very stable. 
We speculate that perhaps an elongated (or otherwise deformed) T1 isomer, where
the degeneracy at the Fermi energy is removed, can reproduce this feature. 
A simple geometry optimization is not enough to investigate this possibility, 
and {\it ab initio} MD simulations will be necessary.

CP clusters with stacking faults are potential candidates for the most stable 
isomer in some occasions, and we have demonstrated this for Al$_{100}$, where 
the energetic difference between the FCC and SF clusters is negligible. 
Therefore, these structural motifs are competitive even at relatively large 
cluster sizes ($N\sim 100$). Experiments indicate that the octahedral FCC 
isomers start to dominate the Al mass-spectrum at $N\ge 200$ due to formation 
of large (111) facets that minimize the surface energy. 
\cite{martin92,martin96}

\section{Acknowledgments}

This work has been supported by the Academy of Finland under the Finnish 
Centre of Excellence Programme 2000-2005 (Project No. 44875, Nuclear and 
Condensed Matter Programme at JYFL). J.A. has been supported by the 
Bundesministerium f\"ur Bildung und Forschung (BMBF), Bonn, within the 
Kompetenzzentrum Materialsimulation, 03N6015. We thank L.-S. Wang
for providing us the experimental PES of Al$_{100}^-$, and R.O. Jones
for a critical reading of the manuscript.


\begin{thebibliography}{30}
\bibitem{tvib} The difference in total energy is converted to vibrational 
temperature according to the classical statistical 
mechanics by taking into account the $3N-6$ degrees of freedom.
\end{thebibliography}
\end{document}